\documentclass[twocolumn,english,conference]{IEEEtran}

\usepackage[cmex10]{amsmath}
\usepackage[margin=0.9in]{geometry}

\usepackage{cite}

\usepackage{latexsym}
\usepackage{color}
\usepackage{graphics}
\usepackage{amssymb}
\usepackage{amsthm}

\usepackage{cite}
\usepackage{array}
\usepackage{graphicx}

\usepackage{url}

\newcommand{\argmax}[1]{{\underset{{#1}}{\mathrm{arg\,max}}}}

\newcommand{\vect}[1]{\mathbf{#1}}

\def\Htran{\mbox{\tiny $\mathrm{H}$}}

\theoremstyle{plain}

\newtheorem{theorem}{Theorem}

\newtheorem{lemma}{Lemma}

\newtheorem{assumption}{Assumption}

\begin{document}

\title{Random Access Protocol for Massive MIMO: Strongest-User Collision Resolution (SUCR)}

\IEEEoverridecommandlockouts

\author{\IEEEauthorblockN{Emil Bj{\"o}rnson\IEEEauthorrefmark{1}, Elisabeth de Carvalho\IEEEauthorrefmark{2}, Erik G. Larsson\IEEEauthorrefmark{1}, Petar Popovski\IEEEauthorrefmark{2}}
\IEEEauthorblockA{\IEEEauthorrefmark{1}{Department of Electrical Engineering (ISY), Link\"{o}ping University, Link\"{o}ping, Sweden}}
\IEEEauthorblockA{\IEEEauthorrefmark{2}{Department of Electronic Systems, Aalborg University, Aalborg, Denmark}}
\thanks{This work was performed partly in the framework of the Danish Council for Independent Research (DFF133500273), the Horizon 2020 project FANTASTIC-5G (ICT-671660), the EU FP7 project MAMMOET (ICT-619086), ELLIIT, and CENIIT. The authors would like to acknowledge the contributions of the colleagues in FANTASTIC-5G and MAMMOET.}%
}

\maketitle

\begin{abstract}
Wireless networks with many antennas at the base stations and multiplexing of many users, known as Massive MIMO systems, are key to handle the rapid growth of data traffic. As the number of users increases, the random access in contemporary networks will be flooded by user collisions. In this paper, we propose a reengineered random access protocol, coined strongest-user collision resolution (SUCR). It exploits the channel hardening feature of Massive MIMO channels to enable each user to detect collisions, determine how strong the contenders' channels are, and only keep transmitting if it has the strongest channel gain. The proposed SUCR protocol can quickly and distributively resolve the vast majority of all pilot collisions.
\end{abstract}

\vspace{-1mm}

\section{Introduction}

\label{sec:intro}

Future cellular networks are facing \emph{massive} numbers of connected user equipments (UEs) that jointly request  massive data volumes \cite{Boccardi2014a}. Since the cellular frequency resources are scarce, this calls for orders-of-magnitude improvements in the spectral efficiency (SE). The Massive MIMO (multiple-input multiple-output) network topology, proposed in \cite{Marzetta2010a}, can bring such extraordinary improvements \cite{Bjornson2016a}. The basic idea is to deploy base stations (BSs) with hundreds of antennas and use these to multiplex tens of UEs at the same time-frequency resource. Massive MIMO is primarily for time-division duplex (TDD) systems, where channel reciprocity can be exploited for scalable pilot-based channel estimation. The achievable uplink (UL) and downlink (DL) data throughputs have been analyzed extensively in recent years; for example, in\cite{Marzetta2010a,Jose2011b,Huh2012a,Hoydis2013a,Ngo2013a,Bjornson2016a}.
In contrast, the network access functionality has received little attention in Massive MIMO \cite{Bjornson2015d}, despite the fact that the massive numbers of UEs with intermittent activity require efficient and scalable solutions. 

Random access has a special role in Massive MIMO systems. In the original Massive MIMO concept \cite{Marzetta2010a}, all UEs within a cell use dedicated orthogonal pilot sequences, while the necessary reuse of pilots across cells leads to \emph{inter-cell pilot contamination}. In future crowded scenarios, the number of UEs residing in a cell is also much larger than the number of pilot sequences, thus the pilots cannot be pre-associated with UEs but need to be opportunistically allocated and deallocated to follow their intermittent activity. The papers \cite{Bjornson2015d} and \cite{Sorensen2014a}  investigate scenarios where the UEs send data packages with randomly selected pilot sequences, with the risk for intra-cell pilot collisions/contamination. In this paper, we propose a new random access protocol that can resolve pilot collisions in Massive MIMO before the data transmission begins. 

\vspace{-1mm}

\subsection{Random Access Protocol in LTE}

\vspace{-1mm}

\begin{figure}[t!]
\begin{center}
\includegraphics[width=0.8\columnwidth]{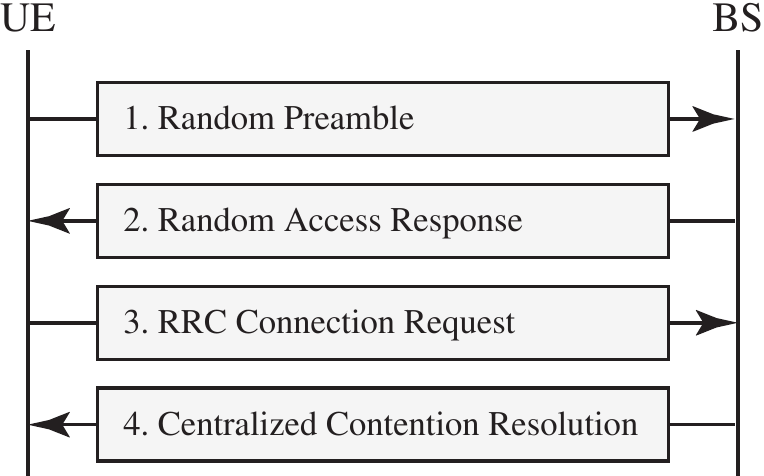}
\end{center} \vskip-5mm
\caption{The PRACH protocol of the LTE system.} \label{figure:PRACH}  \vskip-5mm
\end{figure}

To put our protocol into context, we first review the protocol used on the Physical Random Access Channel (PRACH) in LTE, summarized in Fig.~\ref{figure:PRACH}. In Step~1, the accessing UE picks randomly a preamble from a predefined set. The preamble is a robust entity that enables the BS to gain synchronization. It does not carry a specific reservation information or data and thus has a role as a pilot sequence. Since multiple UEs pick preambles in an uncoordinated way, a collision occurs if two or more UEs select the same preamble. However, at this stage the BS only detects if a specific preamble is active or not \cite{CodeExpanded}. In Step 2, the BS sends a \emph{random access response} corresponding to each activated preamble, to convey physical parameters (e.g., timing advance) and allocate a resource to the UE (or UEs) that activated the preamble.

In Step 3, each UE that has received a response to its transmitted preamble sends a \emph{RRC (Radio Resource Control) Connection Request} in order to obtain resources for the subsequent data transmission. If more than one UE activated that preamble, then all UEs use the same resource to send their RRC connection request in Step 3 and this collision is detected by the BS. Step 4 is called \emph{contention resolution} and contains one or multiple steps that are intended to resolve the collision. This is a complicated procedure that can cause considerable delays.

\vspace{-1mm}

\subsection{Proposed SUCR Random Access Protocol}
\label{subsec:proposed-protocol}

\vspace{-1mm}

The key contribution of this paper is the new strongest-user collision resolution (SUCR) protocol, which reengineers random access for beyond-LTE systems by exploiting the channel hardening feature of Massive MIMO channels. The SUCR protocol consists of four main steps, as illustrated in Fig.~\ref{figure:proposed-protocol}. There is also a preliminary Step 0, in which the BS broadcasts a synchronization signal from which each UE can estimate its average channel gain to the BS. In Step 1, a UE randomly selects a pilot sequence from a predefined set of pilots. This resembles the selection of preambles in LTE, as the collision of two or more UEs that select the same pilot sequence cannot be detected in Step 1. The BS detects which pilot sequences that were used and estimates the channel that each pilot has propagated over. 
If a collision has occurred this becomes an estimate of the superposition of the multiple UE channels.

In Step 2, the BS sends downlink pilot signals that are precoded using the channel estimates. This enables each UE to estimate the sum of the channel gains of the UEs that picked the same pilot and compare it with its own channel gain obtained in Step 0. Each UE can then detect if there has been a collision in Step 1 in a \emph{distributed way}. This departs from the conventional approach in which collisions are detected in a \emph{centralized way} at the BS and broadcasted to the UEs. Based on the detection in Step~2, the UEs can resolve contentions already in Step~3, by applying the local decision rule that only the UE with the strongest channel gain is allowed to retransmit its pilot. This is a key advantage over PRACH in LTE, where all UEs retransmit their preambles in Step~3. Hence, in the SUCR protocol the probability of successful transmission in Step~3 is increased. Step~3 in our protocol resembles the RRC Connection Request; that is, the UE informs about its identity and requests resources to transmit payload data. Step 4 grants these resources or starts a contention resolution, if a collision is detected in Step 3.

Section \ref{sec:system} analyzes the new SUCR protocol in detail, focused on uncorrelated Rayleigh fading channels. The ability to resolve contentions are demonstrated numerically in Section \ref{section:numerical-results}, along with various performance tradeoffs.

\section{System Model and \\ Distributed Pilot Contention Resolution}

\label{sec:system}

We consider a single-cell TDD Massive MIMO system. The time-frequency resources are divided into coherence blocks,  dimensioned such that the channel responses between the BS and the UEs are constant and flat-fading. These coherence blocks are divided into two categories. Most of the blocks are used for coherent UL/DL payload data transmission to active UEs, which have temporarily been allocated dedicated pilot sequences in order to obtain high and reliable data rates. The remaining blocks are dedicated for random access from UEs that wish to obtain dedicated pilot sequences and thereby become active in the payload blocks. The former category can be operated as in the classical works on Massive MIMO \cite{Marzetta2010a,Jose2011b,Huh2012a,Hoydis2013a,Ngo2013a,Bjornson2016a}, while the second category is the main focus of this paper. Random access protocols enable UEs to access the network with low delay whenever needed and to stay active until they run out of data. The network can contain any number of inactive UEs since the BS only allocates dedicated pilots to active UEs and reclaims the pilots when needed. Note that most data applications create intermittent UE activity; even those that are continuous at the application layer.

\begin{figure}[t!]
\begin{center}
\includegraphics[width=0.8\columnwidth]{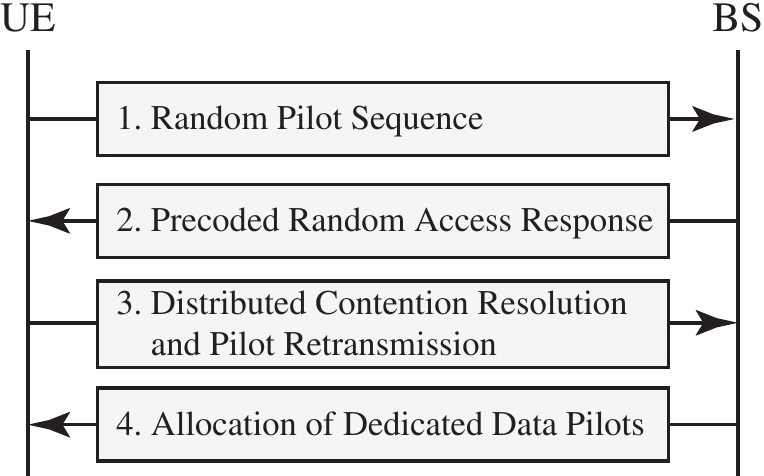}
\end{center} \vskip-5mm
\caption{The proposed SUCR random access protocol for Massive MIMO.} \label{figure:proposed-protocol} \vskip-4mm
\end{figure}

The BS has $M$ antennas and $K$ denotes the number of previously inactive single-antenna UEs that would like to access the network.\footnote{In addition to the $K$ UEs that wish to be active, there can be any number of ``sleeping'' UEs that currently do not attempt to become active.} The random access coherence blocks contain $\tau_p$ orthogonal UL pilots. We typically have $\tau_p \ll K$, but there is no formal constraint. The SUCR protocol consists of four steps, as described in Section~\ref{subsec:proposed-protocol} and Fig.~\ref{figure:proposed-protocol}. It can be successfully applied to most\footnote{Channels with $\vect{h}_k^H \vect{h}_i / M \rightarrow 0$ as $M\rightarrow \infty$, for $i \neq k$, are required.} channels, but we focus on channel responses $\vect{h}_k \in \mathbb{C}^M$ between UE $k$ and the BS that are uncorrelated Rayleigh fading: \vspace{-1mm}
\begin{equation}
\vect{h}_k \sim \mathcal{CN}(\vect{0},\beta_k \vect{I}_M ),
\end{equation} \vskip-1mm
\noindent where the channel gain $\beta_k>0$ describes the path loss.

\begin{figure}[t!]
\begin{center} \vskip+2mm
\includegraphics[width=\columnwidth]{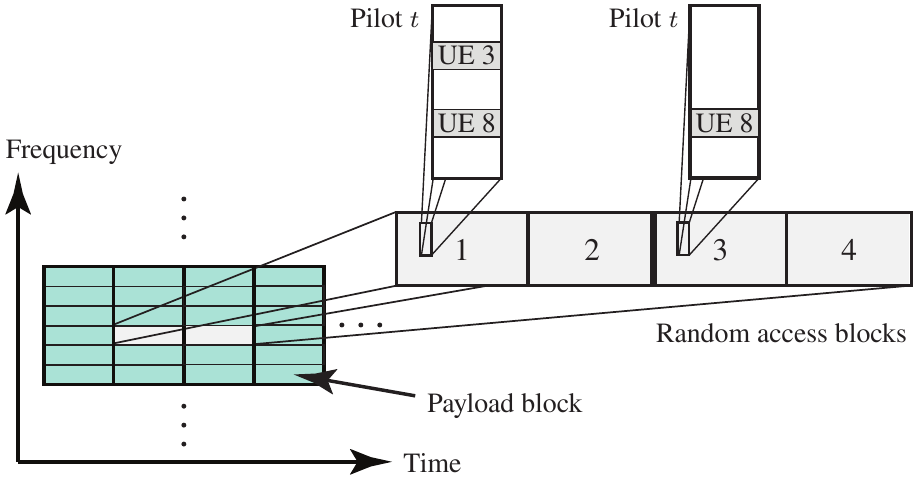}
\end{center} \vskip-3mm
\caption{Example of a transmission protocol where the majority of coherence blocks are used for payload data (where active UEs have been allocated dedicated pilots) and a few for random access (where UEs that wish to become active select pilots at random). Pilot collisions that occur in Step 1 of the random access blocks can be resolved in Step 3.} \label{figure:block-structure} \vskip-3mm
\end{figure}

In what follows, we describe the four steps in the proposed protocol. Fig.~\ref{figure:block-structure} shows an example where these steps are implemented over two coherence blocks.

\vspace{-1mm}

\subsection{Step 1: Random Pilot Sequence}

\vspace{-0.5mm}

Each of the $K$ UEs tries to access the network in Step~1 with probability $P_a \leq 1$, which is a parameter that can be optimized based on the network load; see Section \ref{section:numerical-results}. An access attempt consists of selecting one of the $\tau_p$ pilot sequences in the random access block uniformly at random and transmitting this sequence in Step 1 using a predefined total pilot power $\rho$. Hence, each of the UEs will pick a particular pilot sequence with probability $P_a/\tau_p$. If we focus on an arbitrary pilot sequence, the number of UEs, $N$, that picks this pilot takes a binomial distribution: \vspace{-1mm}
\begin{equation} \label{eq:binomial-dist-N}
N \sim B \left( K, \frac{P_a}{\tau_p} \right). 
\end{equation} \vskip-5mm
\newpage

\noindent The probability that a pilot collision occurs ($N\geq 2$) is
\begin{equation}
1 - \left( 1-\frac{P_a}{\tau_p} \right)^K - K \frac{P_a}{\tau_p} \left(1-\frac{P_a}{\tau_p} \right)^{K-1}.
\end{equation} 
where $(1-\frac{P_a}{\tau_p})^K$ is the probability that the pilot is unused ($N=0$) and  
$K \frac{P_a}{\tau_p} (1-\frac{P_a}{\tau_p})^{K-1}$ that $N=1$ UE uses it.

These pilot collisions need to be detected and the contentions need to be resolved before any UE can be admitted into the payload blocks. We provide a distributed method to resolve pilot collisions at the UE side by utilizing the channel hardening  property of Massive MIMO channels. Without loss of generality, we focus on an arbitrary pilot sequence in a random access block. Let $\mathcal{S} \subset \{1,\ldots,K\}$ be the subset of UEs that have selected this pilot sequence, where 
the cardinality $|\mathcal{S}|=N$ is distributed as in \eqref{eq:binomial-dist-N}. 

The BS receives the signal $\vect{y} \in \mathbb{C}^M$ from the considered pilot sequence in Step 1, where
\begin{equation}
\vect{y} = \sum_{i \in \mathcal{S}} \sqrt{\rho} \vect{h}_i + \vect{n}
\end{equation}
and $\vect{n} \sim \mathcal{CN}(\vect{0}, \sigma^2 \vect{I}_M )$ is additive receiver noise. Note that $\vect{y}$ is obtained by taking the $\tau_p$ received pilot symbols and correlating it with the considered pilot sequence. Hence, $\rho$ represents the total pilot power and is generally $\tau_p$ times larger than the power per transmission symbol.

Since the BS does not know which UEs that have transmitted the pilot, it cannot utilize prior information of the channel gains when estimating the channels. However, it can compute a least-square (LS) estimate of $ \sum_{k \in \mathcal{S}} \vect{h}_k$ as
\begin{equation} \label{eq:channel-estimate-LS}
\hat{\vect{h}}_{\mathrm{LS}} = \frac{1}{\sqrt{\rho}} \vect{y}.
\end{equation}
This estimate is used to determine if $| \mathcal{S} | \geq 1$ or $| \mathcal{S} | = 0$ for the considered pilot sequence; that is, whether or not there is at least one active UE. This can be achieved by computing $\| \hat{\vect{h}}_{\mathrm{LS}}  \|^2 / M$ and compare it to a threshold, which is particularly easy in Massive MIMO systems since
\begin{equation}
\frac{\| \hat{\vect{h}}_{\mathrm{LS}}  \|^2}{M} \rightarrow  \alpha_{\mathcal{S}}  + \frac{\sigma^2}{\rho} \quad \text{as} \quad M \rightarrow \infty,
\end{equation}
due to the law of large numbers, where we have defined
\begin{equation}
\alpha_{\mathcal{S}} = \sum_{i \in \mathcal{S}} \beta_i
\end{equation}
as the sum of the channel gains of the UEs in $\mathcal{S}$. Since a basic criterion for coverage is that the BS can separate a UE signal from the noise, the BS will by definition be able to identify if $| \mathcal{S}|=0$ without errors.
In the remainder of this section we focus on pilots with UEs: $| \mathcal{S} | \geq 1$.

\subsection{Step 2: Precoded Random Access Response}

In Step 2, the BS responds to the pilot signaling by sending orthogonal precoded DL pilot signals that correspond to each of the pilots that were used in the UL. The channel estimate in \eqref{eq:channel-estimate-LS} is used to form a  precoding vector
\begin{equation} \label{eq:precoding-vector}
\vect{w} = \sqrt{q} \frac{\hat{\vect{h}}_{\mathrm{LS}} }{\| \hat{\vect{h}}_{\mathrm{LS}}  \|}
\end{equation}
where the DL total pilot power $q$ has a predefined value. Note that $\vect{w}$ corresponds to maximum ratio transmission (MRT) for the superposition of the UE channels.

By using the precoding vector in \eqref{eq:precoding-vector}, the received DL pilot signal $z_k \in \mathbb{C}$ at UE $k \in \mathcal{S}$ is
\begin{equation} \label{eq:received-user-k}
z_k = \vect{h}_k^{\Htran} \vect{w} + \eta_k
\end{equation}
where $\eta_k \sim \mathcal{CN}(0,\sigma^2)$ is additive receiver noise.
The random distribution of $z_k$ is given by the following lemma.

\begin{lemma} \label{lemma:distribution_z_k}
For any UE $k \in \mathcal{S}$ the received signal can be expressed as $z_k = g_k + \nu_k$, where
\begin{align}
g_k &= \sqrt{ \frac{1}{2} \frac{\rho q \beta_k^2}{\rho \alpha_{\mathcal{S}}  + \sigma^2} } x, \quad x \sim \chi_{2M} \\
\nu_k &\sim \mathcal{CN} \left( 0, \left( \sigma^2 + q \beta_k - \frac{\rho q\beta_k^2}{\rho \alpha_{\mathcal{S}}  + \sigma^2}   \right) \right)
\end{align}
are independent and $\chi_{n}$ denotes the chi-distribution with $n$ degrees of freedom.
\end{lemma}
\begin{IEEEproof}
The proof is given in the appendix.
\end{IEEEproof}

By using the statistics in Lemma \ref{lemma:distribution_z_k}, we notice that the mean and variance of the normalized signal $z_k/ \sqrt{M}$ are
\begin{align} \label{eq:mean-zk/sqrtM}
\mathbb{E}\left\{ \frac{z_k}{\sqrt{M}} \right\}  &= \sqrt{ \frac{\rho q \beta_k^2}{\rho \alpha_{\mathcal{S}}  + \sigma^2} } \frac{\Gamma \left( M + \frac{1}{2} \right) }{ \sqrt{M} \Gamma \left( M \right) } , \\
\mathbb{V} \left\{ \frac{z_k}{\sqrt{M}} \right\}   & = \frac{\rho q \beta_k^2}{\rho \alpha_{\mathcal{S}}  + \sigma^2} \left( 1 - \left( \frac{\Gamma \left( M + \frac{1}{2} \right) }{\sqrt{M} \Gamma \left( M \right) } \right)^2  \right)  \notag \\ &+ \frac{1}{M}\left( \sigma^2 + q \beta_k - \frac{\rho q\beta_k^2}{\rho \alpha_{\mathcal{S}}  + \sigma^2}   \right), \label{eq:variance-zk/sqrtM}
\end{align}
respectively, where $\Gamma(\cdot)$ denotes the gamma function. Furthermore, we notice that
\begin{align}
\mathbb{E}\left\{ \frac{z_k}{\sqrt{M}} \right\} & \rightarrow \sqrt{ \frac{\rho q \beta_k^2}{\rho \alpha_{\mathcal{S}}  + \sigma^2} }, \quad \textrm{as} \quad M \rightarrow \infty, \label{eq:mean-value-z_k-asymptotics} \\
\mathbb{V} \left\{ \frac{z_k}{\sqrt{M}} \right\}  & \rightarrow 0, \quad \textrm{as} \quad M \rightarrow \infty,
\end{align}
by exploiting the fact that  $\frac{\Gamma \left( M + \frac{1}{2} \right) }{ \sqrt{M} \Gamma \left( M \right) } \rightarrow 1$ as $M \rightarrow \infty$. Since the variance approaches zero, the normalized received signal $z_k / \sqrt{M}$ converges to its asymptotic mean in \eqref{eq:mean-value-z_k-asymptotics} as more antennas are added to the BS. This property is often referred to as channel hardening. As a consequence, if the UE knows its own variance $\beta_k$ and the predefined power coefficients ($\rho$ and $q$), it can compute an estimate of $\alpha_{\mathcal{S}}  $ and use it to infer whether or not other UEs have selected the same pilot. The maximum likelihood (ML) estimate of $\alpha_{\mathcal{S}}$ is determined as follows.

\begin{theorem} \label{theorem:ML-estimate-alpha}
The ML estimate of $\alpha_{\mathcal{S}}$ from the observation $z_k = z_{k,\Re} + \jmath z_{k,\Im} $ (with $z_{k,\Re}, z_{k,\Im} \in \mathbb{R}$)  is
\begin{equation}
\hat{\alpha}^{\textrm{ML}}_{\mathcal{S},k} = \argmax{\alpha \geq \beta_k} \quad f_1 \left( z_{k,\Re} | \alpha \right) f_2 \left( z_{k,\Im} | \alpha \right) 
\end{equation}
for the conditional probability density functions (PDFs)
\begin{align}  \label{eq:f1-pdf}
&f_1 \left( z_{k,\Re} | \alpha \right) =  \frac{ e^{- \frac{(z_{k,\Re})^2}{\lambda_2}  \left( 1 - \frac{\lambda_1}{\lambda_1 + \lambda_2}   \right)} }{\Gamma(M) \lambda_1^M \sqrt{\pi \lambda_2 }} \sum_{n=0}^{2M-1} {2M-1 \choose n}   \notag \\
& \times  \frac{\left( \Gamma \left( \frac{n+1}{2} \right) + c_n(z_{k,\Re}) \gamma \left( \frac{n+1}{2} , \frac{(z_{k,\Re})^2}{\lambda_2} \frac{\lambda_1}{\lambda_1+\lambda_2}  \right)  \right) }{   \left( \frac{z_{k,\Re}}{\lambda_2} \right)^{n+1-2M}  \left(  \frac{1}{\lambda_1} + \frac{1}{\lambda_2} \right)^{2M-\frac{n+1}{2}} }  
\\
&f_2 \left( z_{k,\Im} | \alpha \right) = \frac{1}{\sqrt{\pi \lambda_2 }} e^{- \frac{( z_{k,\Im} )^2 }{\lambda_2} } \label{eq:f2-pdf}
\end{align}
where $c_n(z)=(-1)^n\,$ if $z>0$ and $ c_n(z)=-1\,$ if $z<0$, $\gamma(\cdot,\cdot)$ is the lower incomplete gamma function and the coefficients $\lambda_1$ and $\lambda_2$ depend on $\alpha$ as
\begin{align}
\lambda_1 &= \frac{\rho q\beta_k^2}{\rho \alpha  + \sigma^2}    \\
\lambda_2 &= \sigma^2 + q \beta_k - \lambda_1.
\end{align}
\end{theorem}
\begin{IEEEproof}
The proof is given in the appendix.
\end{IEEEproof}

The ML estimate $\hat{\alpha}^{\textrm{ML}}_{\mathcal{S},k}$ can be computed numerically from Theorem \ref{theorem:ML-estimate-alpha}.\footnote{The conditional PDF $f_1 \left( z_{k,\Re} | \alpha \right) $ contains several terms that grow rapidly with $M$, while their ratios remain small. Hence, a careful PDF implementation is needed for numerical stability. The simulations were implemented by taking the logarithm of each term in the summation.} A heuristic estimate in closed-form can be obtained from \eqref{eq:mean-zk/sqrtM} by making the approximation
\begin{equation}
z_{k,\Re} \approx  \mathbb{E}\left\{ z_k \right\}  = \sqrt{ \frac{\rho q \beta_k^2}{\rho \alpha_{\mathcal{S}}  + \sigma^2} } \frac{\Gamma \left( M + \frac{1}{2} \right) }{ \Gamma \left( M \right) } 
\end{equation}
which leads to
\begin{equation} \label{eq:approximate-alpha}
\hat{\alpha}^{\textrm{approx}}_{\mathcal{S},k} = \max \left( \left(   \frac{\Gamma \left( M + \frac{1}{2} \right) }{ \Gamma \left( M \right) }  \right)^2  \frac{q \beta_k^2}{z_{k,\Re}^2} - \frac{\sigma^2}{\rho}, \, \beta_k \right),
\end{equation}
where $\max (\cdot, \cdot)$ takes the maximum of the inputs to make sure that $\hat{\alpha}^{\textrm{approx}}_{\mathcal{S},k} \geq \beta_k$.
The imaginary part is discarded in this approximative estimator since it only contains noise and estimation errors. Note that this approximation is asymptotically tight since the variance of $z_k/\sqrt{M}$ goes to zero as $M \rightarrow \infty$.

\subsection{Step 3: Contention Resolution \& Pilot Retransmission}

The objective of our distributed mechanism for resolving pilot contentions is that each pilot should only be retransmitted by one UE in Step 3. Each UE $k \in \mathcal{S}$ knows its own long-term channel gain $\beta_k$ and has an estimate $\hat{\alpha}_{\mathcal{S},k}$ of the sum of the channel gains of the contending UEs, such that it can infer:
\begin{itemize}
\item If a pilot collision has occurred (i.e., $\hat{\alpha}_{\mathcal{S},k} > \beta_k$);
\item How strong its own channel is relative to the contenders' channels: $\beta_k / \hat{\alpha}_{\mathcal{S},k}$.
\end{itemize}
Since the number of contenders, $| \mathcal{S}|$, is unknown, a UE can only reliably compare its own channel gain with the summation of the gains of its contenders. To resolve the contention we thus make the following assumption.

\begin{assumption}
The contention winner is the UE $k \in \mathcal{S}$ with the largest $\beta_k$, referred to as the strongest user.
\end{assumption}

If $\alpha_{\mathcal{S}}$ would be known, UE $k$ is sure to be the contention winner if $\beta_k > \alpha_{\mathcal{S}} - \beta_k$, irrespective of how many contenders there are. This gives the criterion $\beta_k > \alpha_{\mathcal{S}}/2$.
 We say that we have \emph{resolved a collision} if and only if a single UE appoints itself the contention winner; see Fig.~\ref{figure:block-structure} for an example where a two-UE collision is resolved. This method can always resolve contentions with $|\mathcal{S}|=2$ UEs (and $|\mathcal{S}|=1$ for that matter) under perfect CSI, while there is a risk for \emph{false negatives} for $|\mathcal{S}| \geq 3$ where none of the UEs can identify itself as the contention winner.

Since $\alpha_{\mathcal{S}}$ is estimated in practice, we propose that each UE applies the \emph{activation decision rule}
\begin{align}
&\mathcal{R}_{k}:  \quad \beta_k > \hat{\alpha}_{\mathcal{S},k} /2 + \epsilon_k \quad \textrm{(active)}, \\
&\mathcal{I}_{k}: \quad \beta_k \leq \hat{\alpha}_{\mathcal{S},k} /2 + \epsilon_k \quad \textrm{(inactive)}.
\end{align}
UE $k \in \mathcal{S}$ concludes that it has the largest channel gain if $\mathcal{R}_{k}$ is true and stays active by \emph{retransmitting} the pilot in Step 3. If it instead concludes that $\mathcal{I}_{k}$ is true, it decides to remain \emph{inactive} by pulling out from the random access attempt and try again later. The estimation errors can cause \emph{false positives} where multiple UEs appoint themselves the contention winner. The bias parameter $\epsilon_k \in \mathbb{R}$ is used to tune the system to a performance criterion; for example, maximizing the average number of resolved collisions or minimizing the risk of false positives (or negatives).

The probability of resolving the contention can be characterized based on this decision rule. By numbering the active UEs in $\mathcal{S}$ from $1$ to $N=| \mathcal{S}|$, the probability of resolving the collision is
\begin{equation} \label{eq:prob-resolved}
\begin{split}
P_{N,\textrm{resolved}} &= 
\mathrm{Pr} \{ \mathcal{R}_{1},\mathcal{I}_{2},\ldots,\mathcal{I}_{N} \} \\ &+  
\mathrm{Pr} \{ \mathcal{I}_{1},\mathcal{R}_{2},\mathcal{I}_{3},\ldots,\mathcal{I}_{N} \} + \ldots \\ &+ 
\mathrm{Pr} \{ \mathcal{I}_{1},\ldots,\mathcal{I}_{N-1},\mathcal{R}_{N} \},
\end{split}
\end{equation}
where the randomness is due to channel realizations and noise (and possibly also random user locations).
In the special case of $N=2$, \eqref{eq:prob-resolved} reduces to
\begin{equation} \label{eq:prob-resolved-S2}
\begin{split}
P_{2,\textrm{resolved}} &= 
\mathrm{Pr} \{ \mathcal{R}_{1},\mathcal{I}_{2} \} +  
\mathrm{Pr} \{ \mathcal{I}_{1},\mathcal{R}_{2} \},
\end{split}
\end{equation}
while a false negative occurs if both the UEs pull out (with probability $\mathrm{Pr} \{ \mathcal{I}_{1},\mathcal{I}_{2} \}$) and a false positive occurs when both UEs stay active (with probability $\mathrm{Pr} \{ \mathcal{R}_{1},\mathcal{R}_{2} \}$).
Since $\hat{\alpha}_{\mathcal{S},k} \rightarrow \alpha_{\mathcal{S}}$ as $M \rightarrow \infty$, we can asymptotically resolve all collisions with $|\mathcal{S}|=2$ (as in the perfect CSI case).

\subsection{Step 4: Allocation of Dedicated Payload Pilots}

The BS receives messages from the UEs that remained active in Step 3. These messages can, for example, contain the unique identity number of the UE. If the BS can decode these messages correctly, it can be sure that the potential contentions have been resolved and can admit the corresponding UEs to the payload coherence blocks by allocating temporary dedicated pilot sequences. This resource allocation is broadcasted in Step 4, for example, by using coherent precoding based on the pilots that were sent in Step 3. Remaining contentions can be handled by letting the corresponding UEs try all over again or by initiating further contention resolution as in LTE.

\section{Numerical Results}
\label{section:numerical-results}

In this section, we show numerically that the proposed method can resolve pilot collisions with high probability.
For ease in presentation, we set $\rho=q=\sigma^2=1$ so that $\beta_k$ for the $k$th UE represents the SNR it achieves in the uplink and downlink pilot signaling (per BS antenna).

\begin{figure}[!t]
\begin{center} 
\includegraphics[width=\columnwidth]{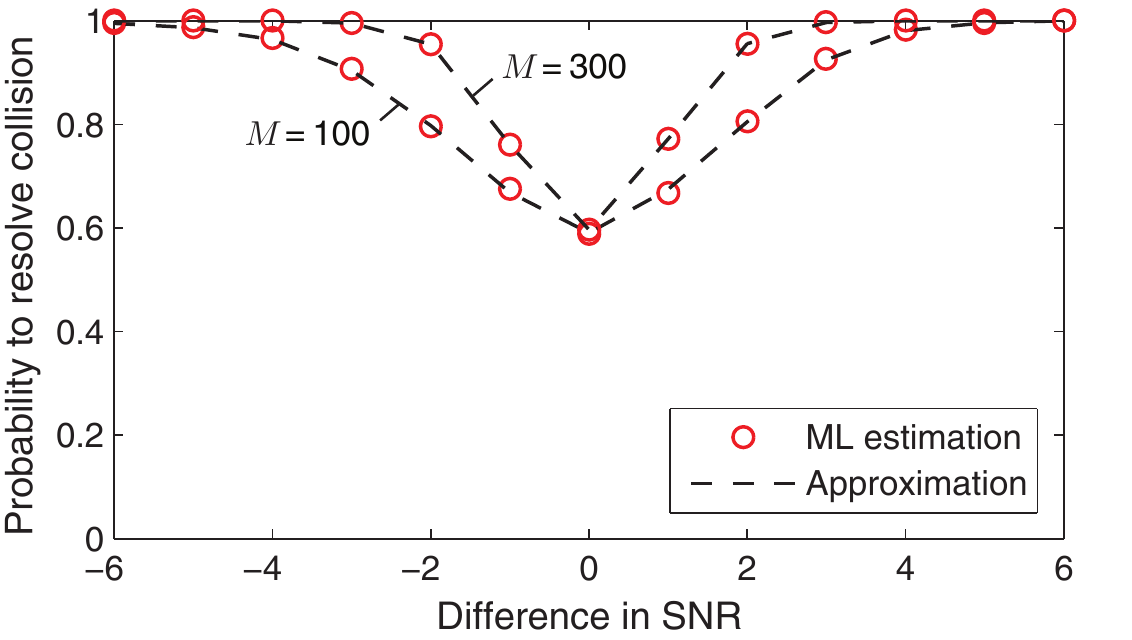}
\end{center}\vskip-3mm
\caption{Probability of resolving a two-UE collision, when UE 1 has $\beta_1=10$ dB and UE 2 has $\beta_2$ between $4$ dB and $16$ dB. ML and approximated estimation of $\alpha_{\mathcal{S}}$ are compared for $M \in \{100,300\}$.} \label{figureAlphaMLNew} \vskip-2mm
\end{figure}

First, we focus on a case where two UEs collide: $\mathcal{S}=\{1,2\}$. The first UE has the fixed SNR $\beta_1 = 10$ dB, while the SNR of the second UE is varied between $4$ dB and $16$ dB. The probability of resolving the collision  $P_{2,\textrm{resolved}} $, as defined in \eqref{eq:prob-resolved-S2}, is shown in Fig.~\ref{figureAlphaMLNew} for $\epsilon_k = 0$ and either $M=100$ or $M=300$ BS antennas. The horizontal axis shows the SNR difference between the UEs, which is between $-6$ dB and $+6$ dB. We compare the result when using the ML estimator from Theorem \ref{theorem:ML-estimate-alpha} with the approximate estimator from \eqref{eq:approximate-alpha}. The curves coincide, thus showing that the approximation is very tight. As expected, the proposed SUCR protocol resolves almost all collisions when $\beta_1$ and $\beta_2$ are sufficiently different (e.g., more than $90\%$ of the two-UE collisions when there is a 3 dB difference in SNR). We notice that adding more antennas improves the probability of resolving the contention, except in the special case $\beta_1 = \beta_2$. This seldom happens since UEs transmit at full power (in contrast to the inversion power control in LTE). Interestingly, the probability of resolving such worst-case collisions is greater than $50\%$, because the estimates $\hat{\alpha}_{\mathcal{S},1} $ and $\hat{\alpha}_{\mathcal{S},2}$ are correlated in the sense that the UE with the most favorable small-scale channel realization obtains a larger estimate.

\begin{figure}[!t]
\begin{center} 
\includegraphics[width=\columnwidth]{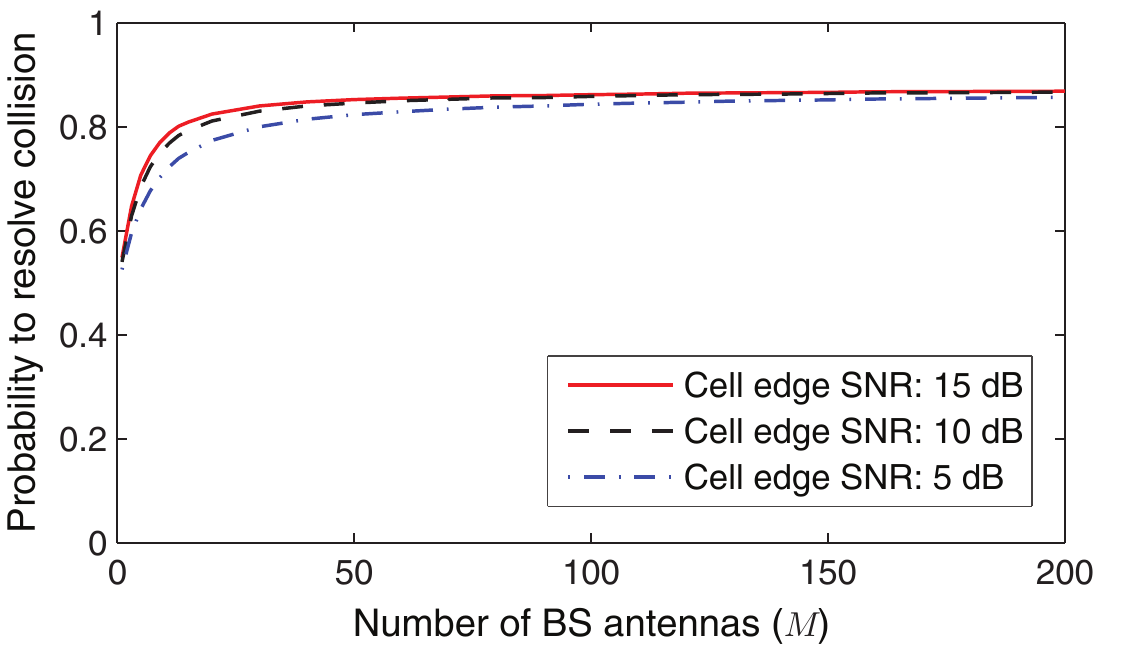}
\end{center}\vskip-4mm
\caption{Probability of resolving collisions, as a function of the number of BS antennas, in a circular cell with different cell-edge SNRs.} \label{figureCircular} \vskip-2mm
\end{figure}

Next, we consider a cellular scenario with $K=50$ UEs uniformly distributed within a circle around the BS with radius $r$, with a minimum distance of $0.1r$. The channel gains are distance-dependent with a pathloss exponent of $3.7$ and we consider log-normal shadow fading with 8 dB standard deviation. Fig.~\ref{figureCircular} shows the probability to resolve collisions, defined based on \eqref{eq:binomial-dist-N} and \eqref{eq:prob-resolved} as
\begin{equation}
\begin{split}
&P_{\textrm{resolved}} = \mathbb{E}\{ P_{N,\textrm{resolved}}  \} \\
&= \sum_{N=1}^{K} P_{N,\textrm{resolved}} {K \choose N}
\left( \frac{P_a}{\tau_p} \right)^N \! \left(1-\frac{P_a}{\tau_p} \right)^{K-N} \!\!,
\end{split}
\end{equation}
as a function of the number of BS antennas. The results based on the approximate estimator of $\alpha_{\mathcal{S}}$, $\tau_p = 10$, $\epsilon_k = 0$, and different cell edge SNRs (defined as the cell-edge $\beta$-value without shadow fading). The results are optimized numerically with respect to the access probability $P_a$.

The first observation is that the proposed SUCR protocol relies on channel hardening; $P_{\textrm{resolved}} $ is around $50\%$ for $M=1$, but increases steeply to around $85\%$ when having $M=50$ antennas. The probability to resolve collisions continues to increase with $M$ for $M \geq 50$, but at a much slower pace. The optimized access probability is then $62\%$, which corresponds to an average of three UEs that selects each pilot. The cell-edge SNR has a clear impact on $P_{\textrm{resolved}} $, where higher SNR leads to more reliable estimation of $\alpha_{\mathcal{S}}$ and thereby more accurate decisions. However, the gain from increasing the SNR saturates at around 10 dB. We stress that a 10 dB cell edge SNR is not large when it comes to pilot signaling, since $\tau_p=10$ implies that the SNR per symbol is only  0 dB.

\begin{figure}[!t]
\begin{center} 
\includegraphics[width=\columnwidth]{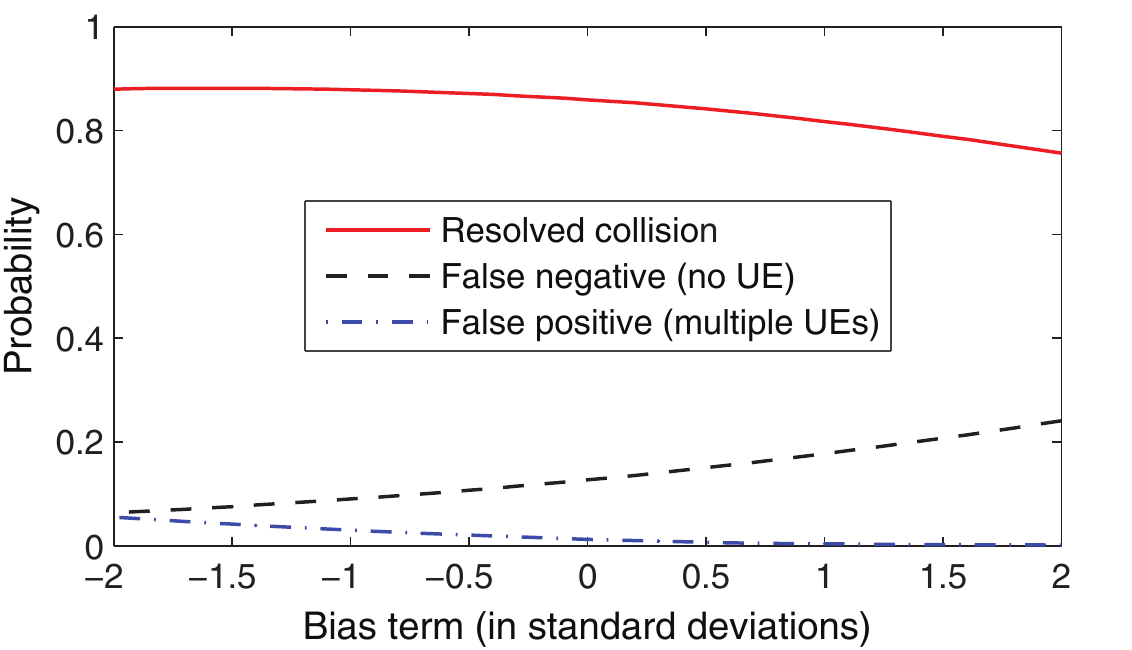}
\end{center}\vskip-4mm
\caption{Probability of resolving collisions, false negatives, and false positives for different bias terms applied in a circular cell.} \label{figureCircularBias} \vskip-3mm
\end{figure}

Finally, we demonstrate how the bias terms $\epsilon_k$ can be utilized to tune the decisions. We consider the same circular cell as in the previous figure, with $M=100$, 10 dB SNR, and $P_a$ optimized for each bias term. Fig.~\ref{figureCircularBias} shows the probability of resolving collisions, false negatives, and false positives for bias terms of the type $\epsilon_k = \delta \beta_k/ \sqrt{M}$, which corresponds to $\delta$ standard deviations of $\| \vect{h}_k \|^2 / M$ around its mean value $\beta_k$.
By subtracting one or two standard deviations from $\hat{\alpha}_{\mathcal{S},k} /2$ in the decision rule, we can encourage UE $k$ to appoint itself the contention winner. This leads to higher probability of resolving collisions, at the cost of more false positives where multiple UEs are still active in Step 3. In contrast, by adding one or two standard deviations to $\hat{\alpha}_{\mathcal{S},k} /2$ in the decision rule, we can discourage UE $k$ to appoint itself the contention winner and thereby bring the probability of false positives towards zero---at the cost of resolving fewer collisions and having more false negative where no UEs remain in Step 3.

\section{Conclusion}

We have proposed the SUCR random access protocol for beyond-LTE Massive MIMO systems. It resolves the pilot collisions that occurs when UEs select pilot sequences at random in an attempt to access the network, before being allocated dedicated pilots. The SUCR protocol exploits channel hardening to make distributed detection and contention resolution decisions at the UEs, such that each pilot is given to the contender with the strongest channel gain. Simulations show that the proposed protocol can make sure that $90\%$ of the pilots only have a single UE left, while there are respectively $5\%$ false positives and negatives. The bias terms in the protocol can also be selected to get virtually no false positives (i.e., no pilots with multiple remaining UEs), at the cost of reducing the average number of single-UE pilots to $75\%$. This removes the need for the cumbersome centralized contention resolution that is used in contemporary systems, such as LTE.

\vspace{-2mm}

\appendix

\textbf{Proof of Lemma \ref{lemma:distribution_z_k}:} If the set $\mathcal{S}$ is known, the MMSE estimator of $ \vect{h}_k$ from the observation $\vect{y}$ is
\begin{equation}
\hat{\vect{h}}_{k,\mathrm{MMSE}} = \frac{\sqrt{\rho} \beta_k}{\rho \alpha_{\mathcal{S}}  + \sigma^2 } \vect{y}.
\end{equation}
This can also be expressed as $\vect{h}_k = \hat{\vect{h}}_{k,\mathrm{MMSE}} + \vect{e}_k$, where $\hat{\vect{h}}_{k,\mathrm{MMSE}} \sim \mathcal{CN}(\vect{0}, \frac{\rho \beta_k^2}{\rho \alpha_{\mathcal{S}}  + \sigma^2} \vect{I}_M )$ is independent from the estimation error $\vect{e}_k \sim \mathcal{CN}(\vect{0}, (\beta_k - \frac{\rho \beta_k^2}{\rho \alpha_{\mathcal{S}}  + \sigma^2} ) \vect{I}_M )$. Notice that  \vskip-5mm
\begin{equation}
\frac{\hat{\vect{h}}_{k,\mathrm{MMSE}} }{\| \hat{\vect{h}}_{k,\mathrm{MMSE}}  \|} = \frac{\vect{y}}{\| \vect{y} \|} = \frac{\hat{\vect{h}}_{\mathrm{LS}} }{\| \hat{\vect{h}}_{\mathrm{LS}}  \|},
\end{equation}
thus the received signal in \eqref{eq:received-user-k} can be rewritten as
\begin{equation}
\begin{split}
z_k = \underbrace{\sqrt{q} \| \hat{\vect{h}}_{k,\mathrm{MMSE}}  \|}_{=g_k} +  \underbrace{\sqrt{q} \frac{\vect{e}_k^{\Htran} \hat{\vect{h}}_{k,\mathrm{MMSE}} }{\| \hat{\vect{h}}_{k,\mathrm{MMSE}}  \|} + \eta_k}_{= \nu_k}.
\end{split}
\end{equation} \vskip-1mm
These factors can be shown to be independent and $\frac{ \vect{e}_k^{\Htran}\hat{\vect{h}}_{k,\mathrm{MMSE}} }{\| \hat{\vect{h}}_{k,\mathrm{MMSE}}  \|} \sim \mathcal{CN}(0, \beta_k - \frac{\rho \beta_k^2}{\rho \alpha_{\mathcal{S}}  + \sigma^2} )$. In addition, $\nu_k$ is the sum of two independent complex Gaussian variables with the variance stated in the lemma. Finally, we notice that $g_k^2$ is the sum of the squares of $2M$ independent Gaussian variables with zero mean and variance $\frac{1}{2} \frac{\rho \beta_k^2}{\rho \alpha_{\mathcal{S}}  + \sigma^2} $, thus $g_k$ has a scaled chi-distribution as stated in the lemma.

\vspace{1mm}

\textbf{Proof of Theorem \ref{theorem:ML-estimate-alpha}:}  The ML estimator is defined as
\begin{equation} \label{eq:MLE}
\hat{\alpha}_{\mathcal{S},k}^\star = \argmax{\alpha} \quad f \left( z_{k,\Re}, z_{k,\Im} | \alpha \right)
\end{equation}
where $f \left( z_{k,\Re}, z_{k,\Im} | \alpha \right)$ is the joint PDF. 
Notice that $z_{k,\Re}=g_k + \Re(\nu_k)$ and $z_{k,\Im}= \Im(\nu_k)$ are independent since $\nu_k \sim \mathcal{CN}(0,\lambda_2)$ has independent real and imaginary parts. 
Hence, $f \left( z_{k,\Re}, z_{k,\Im} | \alpha \right) = f_1 \left( z_{k,\Re} | \alpha \right) f_2 \left( z_{k,\Im} | \alpha \right) $, where $f_2 $ in \eqref{eq:f2-pdf} is the PDF of $\Im(\nu_k)$.  The PDF $f_1 \left( z_{k,\Re} | \alpha \right)$ of $z_{k,\Re}$ is computed as a convolution between the marginal PDFs of $g_k$ and $\Re(\nu_k)$, which eventually yields  \eqref{eq:f1-pdf}.

\bibliographystyle{IEEEbib}
\bibliography{IEEEabrv,refs}

\end{document}